\documentclass[allclo]{FBSart}
\usepackage{amsfonts}
\usepackage{amssymb}
        \usepackage[dvips]{graphicx}
\title{Four-quark stability%
\thanks{Article based on the presentations
by J.~Vijande and J.-M.~Richard at the Fifth Workshop on Critical Stability, Erice, Sicily, Received February 2,  2009; Accepted February 9, 2009.}}

\author{J. Vijande\instnr{1},
A. Valcarce\instnr{2}, 
J.-M. Richard\instnr{3}, 
N. Barnea\instnr{4}}
\instlist{Departamento de F\'{\i}sica At\'{o}mica, Molecular y Nuclear, Universidad de Valencia (UV)  
and IFIC (UV-CSIC), Valencia, Spain.
\and Departamento de F\'\i sica Fundamental, Universidad de Salamanca, 
E-37008 Salamanca, Spain
\and Laboratoire de Physique Subatomique et Cosmologie, Universit\'e Joseph Fourier--INPG--IN2P3-CNRS 53, avenue des Martyrs, 38026  Grenoble, France
\and The Racah Institute of Physics, The Hebrew University, 91904,
Jerusalem, Israel}

\runningauthor{J. Vijande}
\runningtitle{Four--quark stability}
\begin{document}

\maketitle
\begin{abstract}
The physics of charm has become one of the best laboratories exposing the
limitations of the naive constituent quark model and also giving hints
into a more mature description of meson spectroscopy, beyond the simple quark--antiquark configurations. In this talk we review some recent studies of multiquark components in 
the charm sector and discuss in particular
exotic and non-exotic four-quark systems, both with pairwise and many-body forces.
\end{abstract}

More than thirty years after the so-called November revolution~\cite{Bjo85}, 
heavy hadron spectroscopy remains a challenge. The formerly comfortable 
world of heavy mesons is shaken by new 
results~\cite{Ros07}. This  started in 2003 
 with the discovery of the $D_{s0}^*(2317)$ and  $D_{s1}(2460)$ 
 mesons in the open-charm sector. These positive-parity states have masses lighter than expected from quark  models, and
also smaller widths. Out of the many proposed explanations, 
the unquenching of the naive quark model has been successful~\cite{Vij05}. 
When a $(q\bar q)$ pair occurs in a $P$-wave but can couple to hadron 
pairs in $S$-wave, the latter configuration distorts the $(q\bar q)$ picture. 
Therefore, the $0^+$ and $1^+$ $(c\bar s)$ states
predicted above the $DK(D^*K)$ thresholds couple to the continuum. This 
mixes meson--meson components in the wave function,  an idea advocated long 
ago to explain the spectrum and properties 
of light-scalar mesons~\cite{Jaf07}.

This possibility of $(c\bar{s} n\bar{n})$ ($n$ stands for a light quark) components in $D_s^*$ has open the discussion about the presence of compact 
$(c\bar c n\bar n)$ 
four-quark states in the charmonium spectroscopy.  Some states recently found in the hidden-charm sector may fit in the 
simple quark-model description as $(c\bar c)$  pairs 
(e.g., $X(3940)$, $Y(3940)$, and $Z(3940)$ 
as radially excited  $\chi_{c0}$, $\chi_{c1}$, and $\chi_{c2}$), but
 others appear to be more elusive, in particular $X(3872)$, $Z(4430)^+$, and $Y(4260)$. 
The  debate on the nature of these states is open,   with special emphasis on the 
$X(3872)$.
Since it was first reported by Belle in 2003~\cite{Bel03}, 
it has gradually become the flagship of the new armada of states 
whose properties make their identification as
traditional $(c\bar c)$ states unlikely.
An average mass of $3871.2\pm0.5\;$MeV and a narrow width of less than $2.3\;$MeV
have been reported for the $X(3872)$.
Note the vicinity of this state to the $D^0\overline{D}{}^{*0}$ threshold,
$M(D^0\,\overline{D}{}^{*0})=3871.2\pm1.2\;$MeV.
With respect to the $X(3872)$ quantum numbers, although some caution is still 
required until better statistic
is obtained~\cite{Set06}, an isoscalar $J^{PC}=1^{++}$ state seems to
be the best candidate to describe the properties of the $X(3872)$.

Another hot sector, at least for theorists, includes the $(cc\bar n\bar n)$ states, which are manifestly exotic with 
charm $2$ and  baryon number $0$. Should they lie below the threshold for dissociation into two ordinary hadrons, they would be narrow and show up clearly in the experimental 
spectrum. There are already estimates of the production rates
indicating they could be produced and detected at present 
(and future) experimental facilities~\cite{Fab06}.
The stability of such $(QQ\bar q\bar q)$ states has been discussed since the early 80s \cite{Ade82}, and there is a consensus that stability is reached when the mass ratio $M(Q)/m(q)$ becomes large enough.  See, e.g., \cite{Janc} for Refs. This effect is also found in QCD sum rules \cite{Navarra:2007yw}.
This  improved binding when $M/m$ increases is due to the same mechanism by which the hydrogen molecule $(p,p,e^-,e^-)$ is much more bound than the positronium molecule $(e^+,e^+,e^-,e^-)$. What matters is not the Coulomb character of the potential, but its property to remain identical when the masses change. In quark physics, this property is named \emph{flavour independence}.  It is reasonably well satisfied, with departures mainly due to spin-dependent corrections.

The question is whether stability is already possible for $(cc\bar n\bar n)$ or requires heavier quarks. In Ref.~\cite{Janc}, a marginal binding was found for a specific potential for which earlier studies found no binding. This illustrates how difficult are such four-body calculations.

In another recent investigation, the four-body  Schr\"odinger equation
has been solved accurately using the hyperspherical harmonic (HH) formalism~\cite{Vij07}, with 
 two standard quark models
containing a linear confinement supplemented by a Fermi--Breit one-gluon exchange
interaction (BCN), and also boson exchanges
between the light quarks (CQC). The model parameters were tuned
in the meson and baryon spectra. 
The results are given in Table~\ref{t12}, indicating the quantum numbers of the
state studied, the maximum value of the grand angular momentum used in the HH 
expansion, $K_{\rm m}$, and the
energy difference between the mass of the 
four-quark state, $E_{4q}$, and that of the lowest two-meson
threshold calculated with the same potential model, $\Delta_E$. For
the $(cc\bar n \bar n)$ system we have also calculated 
the radius of the four-quark
state, $R_{4q}$, and its ratio to the sum of the radii of the 
lowest two-meson threshold, $\Delta_R$.

%\begin{table}[t]
%\centering
%\caption{ $(c\bar c n\bar n$) results.}
%\begin{tabular}{|c|cc|cc|} 
%\hline
% &\multicolumn{2}{|c|}{CQC} &\multicolumn{2}{|c|}{BCN} \\
%\hline
%$J^{PC}(K_{\rm max})$ & $E_{4q}$ & $\Delta_{E}$&$E_{4q}$ & $\Delta_{E}$ \\ 
%\hline
%\hline
%$0^{++}$ (24) & 3779 &  +34 &  3249 &  +75  \\
%$0^{+-}$ (22) & 4224 &  +64 &  3778 & +140  \\
%$1^{++}$ (20) & 3786 &  +41 &  3808 & +153  \\
%$1^{+-}$ (22) & 3728 &  +45 &  3319 &  +86  \\
%$2^{++}$ (26) & 3774 &  +29 &  3897 &  +23  \\
%$2^{+-}$ (28) & 4214 &  +54 &  4328 &  +32  \\
%$1^{-+}$ (19) & 3829 &  +84 &  3331 & +157  \\
%$1^{--}$ (19) & 3969 &  +97 &  3732 &  +94  \\
%$0^{-+}$ (17) & 3839 &  +94 &  3760 & +105  \\
%$0^{--}$ (17) & 3791 & +108 &  3405 & +172  \\
%$2^{-+}$ (21) & 3820 &  +75 &  3929 &  +55  \\
%$2^{--}$ (21) & 4054 &  +52 &  4092 &  +52 \\
%\hline
%\end{tabular}
%\label{t1}
%\end{table}
%

Besides trying to unravel the possible existence of
bound $(cc\bar n\bar n)$ and $(c\bar c n\bar n)$ states
one should aspire to understand whether it is possible to differentiate between compact
and molecular states. A molecular state may be understood as a four-quark state
containing a single physical two-meson component, i.e., a unique singlet-singlet
component in the colour wave function with well-defined spin and isospin quantum numbers.
One could expect these states not being deeply bound and therefore having a size
of the order of the two-meson system, i.e., 
$\Delta_R\sim1$. Opposite to that,
a compact state may be characterized by its involved structure on the
colour space, its wave function containing different singlet-singlet
components with non negligible probabilities. One would expect
such states would be smaller than typical two-meson systems, i.e.,
$\Delta_R < 1$. Let us notice that while $\Delta_R>1$ but finite would 
correspond to a meson-meson molecule 
$\Delta_R\stackrel{K\to\infty}{\longrightarrow}\infty$ 
would represent an unbound threshold.

%\begin{table}[t]
%\centering
%\caption{ $(cc\bar n\bar n)$ results.}
%\begin{tabular}{|c|c|cccc|} 
%\hline
% & &\multicolumn{4}{|c|}{CQC} \\
%\hline
%& $J^{P}(K_{\rm max})$ & $E_{4q}$ & $\Delta_{E}$ &$R_{4q}$ & $\Delta_R$ \\
%\hline
%\hline
% & $0^{+}$ (28) & 4441 &  +15 &  0.624 & $> 1$ \\
% & $1^{+}$ (24) & 3861 &$-$76 &  0.367 & 0.808 \\
%$I=0$ & $2^{+}$ (30) & 4526 &  +27 &  0.987 & $> 1$ \\
% & $0^{-}$ (21) & 3996 &  +59 &  0.739 & $> 1$ \\
% & $1^{-}$ (21) & 3938 &  +66 &  0.726 & $> 1$ \\
% & $2^{-}$ (21) & 4052 &  +50 &  0.817 & $> 1$ \\
%\hline
% & $0^{+}$ (28) & 3905 &  +50 &  0.817 & $> 1$ \\
% & $1^{+}$ (24) & 3972 &  +33 &  0.752 & $> 1$  \\
%$I=1$ & $2^{+}$ (30) & 4025 &  +22 &  0.879 & $> 1$ \\
% & $0^{-}$ (21) & 4004 &  +67 &  0.814 & $> 1$ \\
% & $1^{-}$ (21) & 4427 &  +1  &  0.516 & 0.876 \\
% & $2^{-}$ (21) & 4461 &$-$38 &  0.465 & 0.766 \\
%\hline
%\end{tabular}
%\label{t2}
%\end{table}

\begin{table}[t]
\caption{\label{t12} $(c\bar{c}n\bar{n})$ (left) and $(cc\bar n\bar n)$ (right) results.}
\centerline{
\begin{tabular}{|ccccc|} 
\hline
 $(c\bar{c}n\bar{n})$&\multicolumn{2}{c}{CQC} &\multicolumn{2}{c|}{BCN} \\
\hline
$J^{PC}(K_{\rm m})$ & $E_{4q}$ & $\Delta_{E}$&$E_{4q}$ & $\Delta_{E}$ \\ 
\hline
\hline
$0^{++}$ (24) & 3779 &  +34 &  3249 &  +75  \\
$0^{+-}$ (22) & 4224 &  +64 &  3778 & +140  \\
$1^{++}$ (20) & 3786 &  +41 &  3808 & +153  \\
$1^{+-}$ (22) & 3728 &  +45 &  3319 &  +86  \\
$2^{++}$ (26) & 3774 &  +29 &  3897 &  +23  \\
$2^{+-}$ (28) & 4214 &  +54 &  4328 &  +32  \\
$1^{-+}$ (19) & 3829 &  +84 &  3331 & +157  \\
$1^{--}$ (19) & 3969 &  +97 &  3732 &  +94  \\
$0^{-+}$ (17) & 3839 &  +94 &  3760 & +105  \\
$0^{--}$ (17) & 3791 & +108 &  3405 & +172  \\
$2^{-+}$ (21) & 3820 &  +75 &  3929 &  +55  \\
$2^{--}$ (21) & 4054 &  +52 &  4092 &  +52 \\
\hline
\end{tabular}
\begin{tabular}{|ccccc|} 
\hline
$(cc\bar n\bar n)$&\multicolumn{4}{c|}{CQC} \\
\hline
$IJ^{P}(K_{\rm m})$ & $E_{4q}$ & $\Delta_{E}$ &$R_{4q}$ & $\Delta_R$ \\
\hline
\hline
$0\,0^{+}$ (28) & 4441 &  +15 &  0.624 & $> 1$ \\
$0\,1^{+}$ (24) & 3861 &$-$76 &  0.367 & 0.808 \\
$0\,2^{+}$ (30) & 4526 &  +27 &  0.987 & $> 1$ \\
$0\,0^{-}$ (21) & 3996 &  +59 &  0.739 & $> 1$ \\
$0\,1^{-}$ (21) & 3938 &  +66 &  0.726 & $> 1$ \\
$0\,2^{-}$ (21) & 4052 &  +50 &  0.817 & $> 1$ \\
\hline
$1\,0^{+}$ (28) & 3905 &  +50 &  0.817 & $> 1$ \\
$1\,1^{+}$ (24) & 3972 &  +33 &  0.752 & $> 1$  \\
 $1\,2^{+}$ (30) & 4025 &  +22 &  0.879 & $> 1$ \\
$1\,0^{-}$ (21) & 4004 &  +67 &  0.814 & $> 1$ \\
$1\,1^{-}$ (21) & 4427 &  +1  &  0.516 & 0.876 \\
$1\,2^{-}$ (21) & 4461 &$-$38 &  0.465 & 0.766 \\
\hline
\end{tabular}
}
\end{table}

As can be seen in Table~\ref{t12} (left), in the case of the $(c\bar c n\bar n)$
there appear no bound states for any set of quantum numbers, including
the suggested assignment for the $X(3872)$. 
Independently of the quark--quark interaction and the quantum numbers 
considered, the system evolves to a
well separated two-meson state. This is clearly seen
in the energy, approaching the threshold made of two free mesons, 
and also in the probabilities of the
different colour components of the wave function
and in the radius~\cite{Vij07}. Thus, in any manner one can claim for the existence
of a bound state for the $(c\bar c n \bar n)$ system. 

A completely different behaviour is observed in Table~\ref{t12} (right).
Here, there are some particular quantum numbers 
where the energy is quickly stabilized below
the theoretical threshold.
Of particular interest is the $1^+$ $cc\bar n\bar n$ state, whose existence 
was predicted more than twenty years ago~\cite{Zou86}. There is a remarkable agreement on the 
existence of an isoscalar $J^P=1^+$ $cc\bar n\bar n$ bound state using both BCN and CQC models, if not in its properties. 
For the CQC model the predicted binding energy is large, $-$ 76 MeV, $\Delta_R<1$, and a very involved structure of its wave function (the $DD^*$ component of its wave function only accounts for the 50\% of the total probability) what
would fit into compact state. Opposite to that, the BCN model
predicts a rather small binding, $-$7 MeV, and $\Delta_R$ is larger than 1, although finite. This state
would naturally correspond to a meson-meson molecule.

Concerning the other two states that are below threshold in Table~\ref{t12} a more careful analysis is required.
Two-meson thresholds must be determined assuming quantum number conservation 
within exactly the same scheme used in the four--quark calculation. 
Dealing with strongly interacting particles, the two-meson states should have well defined total angular 
momentum, parity, and a properly symmetrized wave function if two identical mesons
are considered (coupled scheme). 
When noncentral forces are not taken into account, orbital angular 
momentum and total spin are also good quantum numbers (uncoupled scheme). 
We would like to emphasize that although we use central forces in our calculation the coupled scheme is the relevant one
for observations, since a small non-central component in the
potential is enough to produce a sizeable effect on the width of a state. These state are below the thresholds given by the
uncoupled scheme but above the ones given within the coupled scheme 
what discard these quantum numbers as promising candidates for being observed
experimentally.

Binding increases for larger $M/m$, but in the $(bb\bar n\bar n)$ sector, there is no proliferation of bound states.
We have studied all ground states of $(bb\bar n\bar n)$ 
using the same interacting potentials as in the double-charm case. Only four bound states have been found,  with quantum numbers $J^P(I)=1^+(0)$
, $0^+(0)$, $3^-(1)$, and $1^-(0)$. The first three ones correspond to compact states.

Now, one could question the validity of the potential models used in these estimates, or more precisely, of the extrapolation from mesons to baryons, and then to multiquark states. For the short-range terms, in particular one-gluon exchange, the additive rule
\begin{equation}\label{ll}
V=-\frac{3}{16}\sum_{i<j} \tilde\lambda^{(c)}_i. \tilde\lambda^{(c)}_j\,v(r_{ij})~,
\end{equation}
is justified. 
Here $v(r)$ is the quark--antiquark potential governing mesons, and $ \tilde\lambda^{(c)}_i$ is the colour generator. This is the non-Abelian version of the $1/r \rightarrow \sum q_iq_j/r_{ij}$ rule in atomic physics.

The confining part, however, is hardly of pairwise character. Several authors have proposed that the linearly rising potential $\sigma r$ of mesons ($\sigma$ is the string tension) is generalised as
\begin{equation}\label{FT}
V=\sigma\min(d_1+d_2+d_3)~,
\end{equation}
where $d_i$ is the distance from the $i^{\rm th}$ quark to a junction whose location is optimised, exactly as in the famous problem of Fermat and Torricelli. Unfortunately, the potential (\ref{FT}) differs little from the empirical ansatz (\ref{ll}) which here reduces to $\sigma(r_{12}+r_{23}+r_{31})/2$. Hence baryon spectroscopy cannot probe the three-body character of confinement.

\begin{figure}[tb]
\centerline{{\includegraphics[width=.2\textwidth]{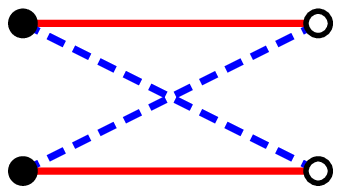}}\hspace*{40pt}
{\includegraphics[width=.2\textwidth]{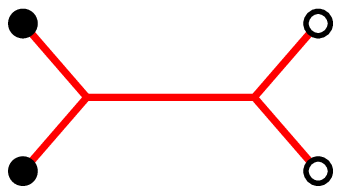}}}
\caption{\label{f1} String model for four quarks: flip-flop (left) and Steiner-tree (right), an alternative configuration that is favoured when the quarks (full disks) are well separated from the antiquarks (open circles).}
\end{figure}

In the case of two quarks and two antiquarks, the confining potential reads
\begin{equation}
V_4=\min(V_f,\,V_s)~,
\end{equation}
 given by the minimum of a flip-flop potential  $V_f$ and a Steiner-tree potential $V_s$, sometimes named ``butterfly'' (see Fig.~\ref{f1}).
 In $V_f$, each gluon flux goes from a quark to an antiquark. The second term corresponds to a  minimal Steiner tree, with four terminals and two Steiner points. It is remarkable that this potential, which is supported by lattice QCD \cite{Okiharu:2004ve} is more attractive than the additive ansatz. This is illustrated in Ref.~\cite{Vij07b}, where the four-body problem is solved with this confining term alone without short-range corrections. The results are displayed in Table \ref{manybody}. This four-body calculation is rather involved, as the potential at each point is obtained by a minimisation over several parameters. See Ref.~\cite{Vij07b} for technical details about the models and the numerical techniques used.

\begin{table}[tb]
\centering
\caption{\label{manybody} Four--quark variational energy $E_4$ of $QQ\overline{q}\overline{q}$ for the different confinement models ($V_f$ stands for the flip-flop interaction, $V_s$ for the Steiner-tree potential, and $V_4=\min(V_f,V_s)$), compared to its threshold, and variational energy $E'_4$ of $Q\overline{Q}q\overline{q}$ with the flip-flop model $V_f$, compared to its threshold $T'_4$ as a function of the mass ratio.}
\begin{tabular}{|c|cccc|cc|}
\hline
$M/m$	& \multicolumn{3}{c}{$E_4$}	& $T_4$ & $E'_4$ & $T'_4$\\ 
	& $V_f$	& $V_s$	& $V_4$	&  & $V_f$  &\\ 
\hline
1	& 4.644	& 5.886	& 4.639	& 4.676  &  4.644 & 4.676 \\
2	& 4.211	& 5.300	& 4.206	& 4.248  & 4.313  & 4.194\\
3	& 4.037	& 5.031	& 4.032	& 4.086  & 4.193  & 3.959 \\
4	& 3.941	& 4.868	& 3.936	& 3.998  & 4.117  & 3.811\\
5	& 3.880	& 4.754	& 3.873	& 3.942  & 4.060  & 3.705\\
\hline
\end{tabular}
\end{table}

The results for the configurations $(QQ\bar{q}\bar{q})$ and $(Q\overline{Q}q\bar{q})$
are shown in Table \ref{manybody} as function of the heavy-to-light mass ratio. Clearly, as $M/m$ increases, a deeper binding is obtained for the flavour-exotic  $(QQ\bar{q}\bar{q})$ system. For the hidden-flavour $(Q\overline{Q}q\bar{q})$, however, the stability deteriorates, becoming unbound for $M/m\gtrsim 1.2$.

More recently, the stability in this model has been demonstrated rigorously in the limit of very large $M/m$. The first step is to show that
\begin{equation} 
V_4/\sigma \le \frac{\sqrt{3}}{2}\left(|\vec{x}|+|\vec{y}|\right)+|\vec{z}|~,
\end{equation}
in terms of the Jacobi variables, $\vec{x}=\vec{r}_2-\vec{r}_1$, 
$\vec{y}=\vec{r}_4-\vec{r}_3$ and $\vec{z}=(\vec{r}_3+\vec{r}_4-\vec{r}_1+\vec{r}_2)/2$, so that the Hamiltonian describing the relative motion is bounded by
\begin{equation}
H_b=\frac{\vec{p}_x^2}{M}+\sigma\frac{\sqrt{3}}{2}|\vec{x}|+
\frac{\vec{p}_y^2}{m}+\sigma\frac{\sqrt{3}}{2}|\vec{y}|+
\frac{\vec{p}_z^2}{4\mu}+\sigma|\vec{z}|~,
\end{equation}
($\mu$ is the quark--antiquark reduced mass), which is \emph{exactly solvable} for its ground state and gives binding for large $M/m$.  Details will be published shortly \cite{Hyam}.

%Finally, our conclusions can be made more general. If we have an $N$-quark
%system described by two-body interactions in such a way that there exists
%a subset of quarks that cannot make up a physical subsystem, then one may expect
%the existence of $N$-quark bound states by means of central two-body potentials.
%If this is not true one will hardly find $N-$quark bound states~\cite{Lip75}.
%For the particular case of the four--quark states, this conclusion is exact if the confinement is described
%by the first  SU(3) Casimir operator, because when the system is split
%into two-mesons the confining contribution from the two isolated mesons
%is the same as in the four-quark system. Many-body interactions as 
%the ones described here favor the binding only in the flavour exotic $(cc\bar n\bar n)$ case and not for the
%$(c\bar c n\bar n)$ one. Another interesting  possibility to explore would be a modification
%of the Hilbert space. If for some reason particular components of the
%four-quark system (diquarks) would be favored against others, the
%system could be compact~\cite{Mai04}. Lattice QCD calculations~\cite{Ale07}
%confirm the phenomenological expectation that QCD dynamics favors the
%formation of good diquarks~\cite{Jaf07}, i.e., in the scalar positive
%parity channel. However, they are large objects whose relevance to hadron
%structure is still under study. 

To conclude, let us stress again the important difference between the two physical systems which have been considered.
While for the $(c\bar c n\bar n)$, there are two allowed physical {\it decay channels},
$(c\bar c)+(n\bar n)$ and $(c\bar n)+(\bar c n)$, for the $(cc\bar n\bar n)$
only one physical system contains 
the possible final states, $(c \bar n)+(c\bar n)$.
Therefore, a $(c \bar c n \bar n)$ four-quark state will hardly present 
bound states, because the system will reorder itself to become the lightest two-meson state, either
$(c\bar c)+(n\bar n)$ or $(c\bar n)+(\bar c n)$. In other words,
if the attraction is provided by the interaction between
particles $i$ and $j$, it does also contribute to the asymptotic
two-meson state. This does not happen
for the $(c c\bar n\bar n)$ if the interaction between, for example,
the two quarks is strongly attractive. 
In this case there is no asymptotic two-meson
state including such attraction, and therefore the system might bind.

Once all possible $(cc\bar n\bar n)$, $(bb\bar n\bar n)$ and $(c\bar cn\bar n)$ quantum numbers have 
been exhausted very few alternatives remain. If additional bound four-quark states or higher configuration are  
experimentally found, then other mechanisms should be at work, for instance based on diquarks \cite{Jaf07,Mai04,Ale07}.
\begin{acknowledge}
This work has been partially funded by the Spanish Ministerio de
Educaci\'on y Ciencia and EU FEDER under Contract No. FPA2007-65748,
by Junta de Castilla y Le\'{o}n under Contract No. SA016A17, and by the
Spanish Consolider-Ingenio 2010 Program CPAN (CSD2007-00042).
\end{acknowledge}

\begin{small}

\end{small}

\begin{thebibliography}{99}
\bibitem{Bjo85} J.D.~Bjorken, The November Revolution: A Theorist Reminisces, 
in: A Collection of Summary Talks in High Energy Physics (ed. J.D.~Bjorken),
p. 229 (World Scientific, New York, 2003).

\bibitem{Ros07} J.L.~Rosner,
		J. Phys. Conf. Ser. {\bf 69}, 012002 (2007).

\bibitem{Vij05}J.~Vijande, F.~Fern\'andez, and A.~Valcarce,
		Phys. Rev. D. {\bf 73}, 034002 (2006).

\bibitem{Jaf07} R.L.~Jaffe,
		Phys. Rept. {\bf 409}, 1 (2005).

\bibitem{Bel03} Belle Collaboration, S.-K.~Choi {\it et al},
                Phys. Rev. Lett. {\bf 91}, 262001 (2003).

\bibitem{Set06} K.K.~Seth, 
		AIP Conf. Proc. {\bf 814}, 13 (2006).


\bibitem{Fab06} A.~del Fabbro, D.~Janc, M.~Rosina, and D.~Treleani,
                Phys. Rev. D {\bf 71}, 014008 (2005).
                
 \bibitem{Ade82} J.~P.~Ader, J.~M.~Richard, and P.~Taxil, 
                Phys. Rev. D {\bf 25}, 2370 (1982); 
                J.~L.~Ballot and J.~M.~Richard,
                Phys. Lett. B {\bf 123}, 449 (1983).



\bibitem{Janc}
  D.~Janc and M.~Rosina,
  %``The T(cc) = D D* molecular state,''
  Few Body Syst.\  {\bf 35} (2004) 175.
%  [arXiv:hep-ph/0405208].
  %%CITATION = FBSYE,35,175;%%  
  
  
   \bibitem{Navarra:2007yw}
  F.~S.~Navarra, M.~Nielsen and S.~H.~Lee,
  %``QCD sum rules study of $QQ-\bar{u}\bar{d}$ mesons,''
  Phys.\ Lett.\  B {\bf 649} (2007) 166.
%  [arXiv:hep-ph/0703071].
  %%CITATION = PHLTA,B649,166;%%     
  
  
\bibitem{Vij07} J.~Vijande, E.~Weissman, N.~Barnea, and A.~Valcarce,
                        Phys. Rev. D {\bf 76}, 094022 (2007).

\bibitem{Zou86} S.~Zouzou, B.~Silvestre-Brac, C.~Gignoux, and J.-M.~Richard,
                Z. Phys. C {\bf 30}, 457 (1986).  
  
  \bibitem{Okiharu:2004ve}
F.~Okiharu, H.~Suganuma and T.~T. Takahashi,
\newblock Phys. Rev. {\bf D72}, 014505 (2005), [hep-lat/0412012].
%%CITATION = HEP-LAT/0412012;%%       

\bibitem{Vij07b} J.~Vijande, A.~Valcarce, and J.-M.~Richard,
                Phys. Rev. D {\bf 76}, 114013 (2007).
                
 \bibitem{Hyam}
 Cafer Ay, J. Hyam Rubinstein, and J.-M. Richard,
 %``Stability of asymmetric tetraquarks in the minimal-path linear potential,''
  arXiv:0901.3022 [math-ph].                

%\bibitem{Lip75} H.J.~Lipkin,
    %                   Phy. Lett. {\bf 58B}, 97 (1975).

\bibitem{Mai04} L.~Maiani {\it et al},
                        Phys. Rev. Lett. {\bf 93}, 212002 (2004).

\bibitem{Ale07} C.~Alexandrou, Ph.~de~Forcrand, and B.~Lucini,
                         Phys. Rev. Lett. {\bf 97}, 222002 (2006).


\end{thebibliography}
\end{document}